\documentclass[12pt]{article}
\usepackage{graphicx}
\usepackage{amsmath}
\usepackage{amsfonts}
\usepackage{amssymb}
\numberwithin{equation}{section}

\begin{document}

\title{Initial boundary value problems for Einstein's field equations and geometric uniqueness}

\author{
Helmut Friedrich\\ 
Max-Planck-Institut f\"ur Gravitationsphysik\\
Am M\"uhlenberg 1\\
14476 Golm, Germany}

\maketitle

\begin{abstract}

{\footnotesize

\noindent
While there exist now formulations of initial boundary value problems for Einstein's field equations which are well posed and preserve constraints and gauge conditions, the question of geometric uniqueness remains unresolved. For two different approaches we discuss how this difficulty arises under general assumptions. So far it is not known whether it can be overcome  without imposing  conditions on the geometry of the boundary. We point out a natural and  important class of initial boundary value problems which may offer  possibilities to arrive at a fully covariant formulation. 

}
\end{abstract}

{\footnotesize

\section{Introduction}
This article is concerned with the general principle of geometric uniqueness in the context of the initial boundary value problem for Einstein's field equations. Because J\"urgen Ehlers 
had always been attracted by problems of principle and he took a certain interest in the  initial boundary value problem (\cite{ehlers:kind}) I devote this article to his memory.

\vspace{.3cm}

The Cauchy problem for Einstein's field equations has a long history (\cite{chrusciel:friedrich}).  It  was put on a solid basis when Choquet-Bruhat 
proved the first  local existence result  (\cite{foures-bruhat}) and its basic principles are now considered as well understood.  In contrast, the systematic and general study of the initial boundary value problem for Einstein's field equations started only recently and there are still basic open problems. There are two reasons for this. 

The PDE theory of initial boundary value problems is technically more involved than that of the Cauchy problem and it has been developed in sufficient generality only in the last few decades (cf. \cite{benzoni-gavage:serre}, \cite{kreiss:lorenz} for references). On the other hand,  there seem to be only a few applications in general relativity whose geometry asks naturally for an analysis in terms of initial boundary value problems.  Examples for this  are the solutions of Anti-de Sitter type. They can be studied by prescribing for Einstein's vacuum field equations with negative cosmological  constant   data  at space-like and null infinity, which is represented in terms of the conformally rescaled and extended space-time by a time-like hypersurface ${\cal J}$, and  by prescribing  Cauchy data on a space-like initial hypersurface ${\cal S}$  which extends to ${\cal J}$.  The study of this problem in \cite{friedrich:AdS} gives  the first analysis of an initial boundary value problem for Einstein's field equations which is general in the sense that no symmetries are required. Remarkably, the problem to be addressed in the following does not occur. 

The main  interest in initial boundary value problems in general relativity stems from numerical relativity. The need to perform numerical calculations on finite grids suggests to introduce artificial time-like hypersurfaces as boundaries of the grid. The general picture is then that the metric $g$  is calculated on a manifold ${\cal M}$ with boundary  ${\cal S} \cup {\cal T}$ and edge $\Sigma = {\cal S} \cap {\cal T}$.
Here ${\cal S}$ and ${\cal T}$ are hypersurfaces which are space- and time-like respectively for $g$  and  intersect in the space-like surface $\Sigma$, the set ${\cal M} \setminus {\cal S}$ is lying in the future of ${\cal S}$. Such a space-time $({\cal M}, g)$ will be referred to as an $ST$-space-time and it will be called an $ST$-vacuum-space-time if it solves Einstein's vacuum field equations $R_{\mu \nu} = 0$.  
In some of the following remarks one will have to assume that $({\cal M}, g)$ satisfies some suitable version 
of `global hyperbolicity'. We will not discuss this any further here.

To simplify the discussion we assume  all fields to be smooth. In the applications $\Sigma$ will usually  be compact but this will not be important in the following. Because hyperbolic problems can be localized we can focus our attention on some neighbourhood of a given point $p \in \Sigma$. We shall mainly be concerned with evolutions local in time that cover some neighbourhood of ${\cal S}$ (which includes  a neighbourhood of $\Sigma$ in ${\cal T}$). Problems arising in long term evolutions  will be addressed only  if they shed  light on the need to control  geometric uniqueness.

Given an $ST$-vacuum-space-time, one may ask {\it which  data need to be given on ${\cal S}$ and ${\cal T}$ to reconstruct  the space-time uniquely by solving the field equations}.  
 A first  answer to this question is given in \cite{friedrich:nagy}, where it is shown that Einstein's field equations admit well-posed initial  boundary value problems. One of the main difficulties overcome in this work is to ensure that gauge conditions and/or constraints are preserved. For reasons discussed below this task is more subtle here than in the case of the Cauchy problem.

In the subsequent years there was a certain activity concerned with `constraint preserving' formulations of initial boundary value problems  in numerical relativity (cf. \cite{alcubierre} for some of the references).  In the present article we shall consider besides  the formulation of \cite{friedrich:nagy} only the one proposed recently  by Kreiss, Reula, 
Sarbach, and Winicour  (cf. (\cite{kreiss:et:al:2009} and references to earlier work given there). It is based on a general and  systematic analysis of well posed  initial boundary value problems for systems of wave equations. Though the application to Einstein's field equations  is not completely analysed yet, the approach appears to be very flexible  and it may offer chances  to clarify a question which I consider a main open problem of the subject.
While being pointed out at various occasions (\cite{friedrich:dpg}, \cite{friedrich:nagy}) it has  largely been  ignored so far. Analysing  in  local evolution problems  the preservation of constraints and  gauge conditions is obviously a  task which needs to be solved before geometric uniqueness can be addressed.

The space-time structures underlying the general theory of relativity are isometry  classes of space-times. A general discussion of boundary value problems of some sort for Einstein's field equations should  thus produce solutions $g_{\mu \nu}$ which, considered as geometric objects, should  not depend on any extraneous structures such as background metrics, frame fields, gauge source functions, gauge conditions, coordinates, etc.  which are introduced to formulate suitable PDE problems.
It should depend only on the `isometry class  of the data'.

In the case of the Cauchy problem this is guaranteed by a {\it geometric uniqueness theorem} which shows that Cauchy data which are isometric in a suitable sense develop into solutions of Einstein's field equations which are unique up to  isometries if the solutions are required to be globally hyperbolic and maximal with this property  (\cite{choquet-bruhat:geroch}, \cite{ringstroem}). In concrete cases it may not be easy to decide whether two Cauchy data sets are isometric but at least there is a well defined concept available.
In the case of the initial boundary value problem the situation is not so clear.

It should perhaps be emphasized that geometric uniqueness is not just a problem of academic interest. As long as it has not been clarified it may not be clear what do  in analytical studies of initial boundary value problems 
if  the gauge  threatens to break down  in the course of a development and requires a redefinition. 
Also, by just comparing their initial boundary data two numerical relativists may not be able to decide, before they start their calculations, whether they can expect to obtain isometric solutions.  As shown below the  comparison may require the knowledge of the complete development in time of their solutions near the boundary and the subsequent calculation of a gauge transformation on the boundary.

 It is the purpose of this article to illustrate the difficulties to arrive at  an appropriate concept of geometric uniqueness  in the context of general initial boundary value problems. Whether they will  find a natural resolution or whether we will have to live with them remains an open question. At the end of the article we discuss a natural and  important class of initial boundary value problems which seem to offer a chance  for a positive answer.

\section{Some observations}

The following {\it flat, linear model problem} is the  prototype of an initial boundary value problem for a hyperbolic equation. In terms of coordinates $x^{\mu}$  on Minkowski space 
in which  $g_{\mu \nu} = \eta_{\mu \nu} \equiv diag(-1, 1, 1, 1)$ we consider
the wave equation
\[
g^{\mu \nu}\,\phi_{,\,\mu \nu} = f \quad \mbox{on} \quad {\cal M} = \{t \ge 0, |x| \le R\},
\]
where $t = x^0$, $R  = const. > 0$, and $|x| = \sqrt{\sum_{a = 1}^3 (x^a)^2}$.
We prescribe 
 initial conditions
\[
\phi = \phi_0, \quad \partial_t\,\phi = \phi_1 \quad \mbox{on} \quad   {\cal S} = \{t = 0, |x| \le R\}, 
\]
and boundary conditions
\begin{equation}
\label{boundary-cond}
B^{\mu}\,\phi_{,\,\mu} = h  \quad \mbox{on} \quad   {\cal T} = \{t \ge 0, |x| = R\},
\end{equation}
which are defined in terms of the vector field
\begin{equation}
\label{M-def}
B = (1 + e)\,T_* + (1 - e)\,N,
\end{equation}
on ${\cal T}$, where $T_*$ denotes the future directed time-like unit vector field $\partial_t$ tangential to ${\cal T}$,  $N$ the outward directed unit normal to ${\cal T}$,  and
\begin{equation}
\label{e-def}
e \in C^{\infty}({\cal T}, \mathbb{R}), \quad \quad |e| \le 1,
\end{equation}
is a given function.
The initial data $\phi_0$, $\phi_1$, the boundary data $h$, and $f$ 
 are given real-valued  functions which are assumed to be smooth.
 
The authors of  \cite{kreiss:et:al:2009} consider more general boundary conditions. They require $|e| < 1$ in (\ref{boundary-cond}) but admit  arbitrary future directed, unit time-like vector fields $T$ tangential to ${\cal T}$ instead of  $T_*$ and allow for right hand sides of  much more general form. The problem posed above suffices  to illustrate the points we wish to discuss.

Clearly, it can only have a solution which is smooth if the data satisfy certain {\it consistency conditions} along the edge
\[ 
\Sigma = \{t = 0, |x| = R\} = {\cal S} \cap {\cal T}.
\]
The wave equation and the  initial data imply on ${\cal S}$ the relations
\begin{equation}
\label{A-consist-cond}
\partial_t^{2\,k}\,\phi= \Delta^k\,\phi_0 + 
\sum_{l = 0}^{k - 1} \Delta^l\,\partial_t^{2(k - 1 + l)} f|_{t = 0}, 
\end{equation}
\[
\partial_t^{2\,k + 1}\phi = \Delta^k\,\phi_1 + 
\sum_{l = 0}^{k - 1} \Delta^l\,\partial_t^{2(k - 1 + l) + 1} f|_{t = 0},
\quad \quad k = 0, 1, 2, \ldots, 
\]
where $\Delta$ denotes the flat Laplacian. The boundary condition implies on $\Sigma$
\begin{equation}
\label{B-consist-cond}
\sum_{j = 0}^k {k \choose j}\left\{
\partial_t^jB^t\,\partial^{k - j + 1}_t\,\phi + 
\sum_{a = 1}^3\partial_t^jB^a\,(\partial^{k - j}_t\,\phi)_{,\,a}\right\}
= \partial^{k}_t\,h,
\quad \quad k = 0, 1, 2, \ldots.
\end{equation}
For our initial boundary value problem to admit smooth solutions, conditions (\ref{A-consist-cond})  and (\ref{B-consist-cond}), read for given function $f$ as conditions on the data $\phi_0$, $\phi_1$, and $h$, must be satisfied on $\Sigma$. We consider two special situations.

\vspace{.2cm}

\noindent
i) The intended applications requires a fixed boundary datum $h$ and thus fixed values of $\partial^k_t \,h$ on $\Sigma$. Inserting the time derivatives of $\phi$ given by  (\ref{A-consist-cond})  in the relations (\ref{B-consist-cond}) then results in a sequence of conditions on the space derivatives
of the data $\phi_0$ and $\phi_1$ on $\Sigma$. Depending on the function $e$ the construction of such data may be fairly complicated.

\vspace{.2cm}

\noindent
ii) The application requires us to use some prescribed  initial data $\phi_0$, $\phi_1$.  Inserting the  time derivatives of $\phi$ given by  (\ref{A-consist-cond})  in (\ref{B-consist-cond}) determines  then  the values of $\partial^k_t\,h$ on $\Sigma$. Finding functions $h$ on ${\cal T}$ whose time derivatives assume  on $\Sigma$  these values  leaves a large freedom for $h$ in the future of $\Sigma$. 
If, for a given function $f$,  this procedure if it is carried out for all possible data $\phi_0$, 
$\phi_1$, all admissible initial and boundary data are obtained.

\vspace{.2cm}

If the conditions above are satisfied our initial boundary value problem admits a unique smooth solution.
Obviously, this result needs to be largely generalized before it can be applied  to Einstein's field equations. Two quite different such applications  will be discussed in the next chapters. 
Here  we use the example above to draw some general conclusions concerning initial boundary value problems for Einstein's vacuum field equations.

To formulate such problems one has to provide in a first step suitable initial data $h_{ab}$, $\kappa_{ab}$ on a space-like initial hypersurface ${\cal S}$ with boundary 
$\Sigma = \partial {\cal S}$ for Einstein's vacuum field equations.  The fields $h_{ab}$ and $\kappa_{ab}$ are supposed to be isometric to the intrinsic metric and the second fundamental form induced by the prospective space-time solution $g$ on the embedded hypersurface  ${\cal S}$.  They  will thus  have to satisfy the constraints implied by the field equations on space-like hypersurfaces.
 There are known now fairly general methods to construct solutions to the vacuum constraints (\cite{bartnik:isenberg}, \cite{holst:nagy:2008}, \cite{maxwell}). 

The consistency conditions to be satisfied on $\Sigma$ may create difficulties though. 
If the intended application requires us to follow the procedure (i), the data will need to satisfy differential relations at any order on $\Sigma$ (the order will be finite if the differentiability requirements on the solutions are relaxed). Such data cannot be constructed by the conformal standard method, in which the problem of solving  the constraints is immediately reduced to one of solving  elliptic equations. 
More recent methods show more flexibility in exploiting  the underdeterminedness of the constraints (\cite{chrusciel:delay:2003}, \cite{corvino}, \cite{corvino:schoen}). But even if they turn out to be useful  in analytical studies of the present problem there may remain challenges for the numerical relativist.

If the intended application allows us to follow the procedure (ii), it turns out that things are easier also in the case of Einstein's field equations. In the formulation discussed in the next chapter there arises no problem at all, in the second formulation this question has apparently not been analysed in detail but it also appears to be simpler.

\vspace{.3cm}

If we want to analyse the freedom to prescribe boundary data for Einstein's field equations 
on the boundary ${\cal T} \sim \mathbb{R}_0^+ \times \Sigma$, the first observation to be made in the model problem above is that the freedom to prescribe data on the boundary is only half as large as the freedom to prescribe initial data. Let us consider 
Einstein's equations as equations of second order for the metric coefficients.  In {\it wave coordinates}, characterized by the conditions  
\begin{equation}
\label{harm-coord}
0 = \Box_g\,x^{\mu} = - \Gamma^{\mu} = - g^{\lambda \rho}\,\Gamma_{\lambda}\,^{\mu}\,_{\rho},
\end{equation}
the field equations take the form of  a system of wave equations (referred to as the `reduced system') for which only ten boundary conditions can be given. This leads to various complications. To get an idea what has to be achieved by the initial and the boundary conditions, we do some simple function counting. 

On the space-like initial hypersurface $S = \{t \equiv x^0 = 0\}$ the Cauchy data  $g_{\mu \nu}$ and $\partial_t g_{\mu \nu}$ have to be given. The functions $g_{0 \nu}$ and  $\partial_t g_{0 \nu}$ are used to remove the gauge freedom and to satisfy the gauge conditions. The remaining metric coefficients can be identified with the metric $h_{ab}$ and the remaining time derivatives are related to the second fundamental form 
$\kappa_{ab} = \kappa^*_{ab} + \frac{1}{3}\,\kappa\,h_{ab}$ with $h$-trace free part $\kappa^*_{ab}$ and mean extrinsic curvature $\kappa$. Three of the six functions $h_{ab}$ can locally be disposed of by removing the freedom to prescribe the coordinates $x^a$ and a further condition is imposed by the  Hamiltonian constraint. Of the six functions $\kappa_{ab}$ the mean extrinsic curvature can be thought of  as fixing the shape of the embedding of ${\cal S}$ into the solution space-time while the trace free part $\kappa^*_{ab}$ is restricted by the three momentum  constraints. In both cases there remains a freedom of two functions which corresponds to the two degrees of freedom of the gravitational field. 

While these considerations gives some insight into the structure and the role  of the data it should be said that they are  in fact very crude. For instance, it is known that a prescribed mean extrinsic curvature (and possibly a space-like surface $\Sigma$) can be used to determine an embedded space-like hypersurface (with boundary  $\Sigma$) in  a space-time which is given (\cite{bartnik:1984}, \cite{bartnik:1988}). 
In an initial value problem the embedding of the  initial hypersurface is constructed, however, simultaneously with the space-time and the roles of the different parts of the data and the constraints  cannot be seperated from each other so clearly. Nevertheless, we shall discuss the boundary data in a similar way  to illustrate what has to be achieved.

The boundary conditions and data must control the evolution of the boundary. In
\cite{friedrich:nagy} this is achieved by prescribing on the boundary ${\cal T}$ its prospective mean extrinsic curvature and on the edge $\Sigma$ a direction tangential to ${\cal T}$ and  transverse to $\Sigma$.  The coordinate $x^3$ is then defined so that  ${\cal T} = \{x^3 = 0\}$. In \cite{kreiss:et:al:2009}  this has not been completely analysed yet.  We note that an understanding of the mechanism which controls the boundary becomes important if the boundary developes a tendency  to form cusps or selfintersections. 

The boundary conditions have to be given such that constraints and/or gauge conditions are preserved. This requirement, which  is taken care of in quite different ways in 
\cite{friedrich:nagy} and \cite{kreiss:et:al:2009} poses considerable difficulties. In an approach based on wave coordinates an obvious choice would be  to require 
\begin{equation} 
\label{gamma=0-on-T}
\Gamma^{\mu} = 0 \quad \mbox{on} \quad  {\cal T}.
\end{equation}
In such an approach equations (\ref{harm-coord}) are implicit in the reduced evolution system. Thus three of the boundary conditions for the  reduced system must be given so that they comprise suitable boundary condition for the wave equations (\ref{harm-coord}) which govern the development of the  coordinates. 
This leaves us with the freedom to specify two conditions which control the two degrees of freedom in the gravitational field.  Because of the vagueness of the concept of `gravitational degrees of freedom'  it is far from obvious how this to be done.
The main difficulty, however, is how these requirements can be met so that they  result in a well-posed initial boundary value problem for the reduced equations. 
The operators acting on the $g_{\mu \nu}$ in (\ref{gamma=0-on-T}) are already fairly complicated and impose severe restrictions on the remaining choices.

We close this discussion by pointing out two basic differences between initial and boundary data.

\vspace{.1cm}

\noindent
i) The choice of the local coordinates on the initial hypersurface ${\cal S}$ is rather arbitrary and of little consequence for the space-time development local in time. In contrast, the gauge along the boundary ${\cal T}$ is tightly related to the evolution process. 
\vspace{.1cm}

\noindent
ii) Together  with the reduced equations  the Cauchy data on ${\cal S}$ give us control on the geometry at all orders on ${\cal S}$. In contrast, only  very little direct information on the geometry of the boundary ${\cal T}$ is provided by the boundary data and conditions. In general, neither the induced metric nor the second fundamental form is available there before the solution has been determined. 

\vspace{.2cm}

The difficulties of controlling the preservation of constraints and gauge conditions which result from these properties have been overcome. There remains, however, the problem of geometric uniqueness.

\section{An  approach  based on the Bianchi equation}

To illustrate the nature  of the difficulties with geometric uniqueness we need to consider the initial  boundary value problem for  Einstein's vacuum equation in some detail.  We concentrate on those features which are specific to the treatment of the boundary.

\subsection{Formulation of the PDE problem}

In the approach  of \cite{friedrich:nagy} the Einstein equations are expressed in terms of the following unknows

\vspace{.1cm}

\noindent
$-$ the coefficients $e^{\mu}\,_k$ of a frame $\{e_k\}_{k= 0, 1, 2, 3}$ 
in suitable coordinates $x^{\mu}$, $\mu = 0, 1, 2, 3$,\\
\hspace*{.4cm}where the frame  satisfies\footnote
{In \cite{friedrich:nagy} the signature $(1, -1, -1, -1)$ is used.
} 
$g(e_j, e_k) = \eta_{jk} = diag(-1, 1, 1, 1)$,

\vspace{.1cm}

\noindent
$-$ the coefficients $\Gamma_k\,^i\,_j$ of the Levi-Civita connection in the frame $e_k$, which satisfy
with\\
\hspace*{.4cm}the covariant derivative operator $\nabla$ defined by $g$
\[
\nabla_k\,e_j \equiv \nabla_{e_k}\,e_j = \Gamma_k\,^i\,_j\,e_i,
\]
\noindent
$-$ a tensor $C^{i}\,_{jkl}$ with the algebraic properties of a conformal Weyl tensor, which is given\\
\hspace*{.4cm}in the frame $e_k$.

\vspace{.1cm}

\noindent
The field equations are represented by the first structural equation
\begin{equation}
\label{first-structural}
[e_i,e_j] = (\Gamma_i\,^k\,_j - \Gamma_j\,^k\,_i)\,e_k,  
\end{equation}
the second structural equation with the assumption that the Ricci tensor vanishes
\begin{equation}
\label{second-structural}
e_{[k}(\Gamma_{l]}\,^i\,_j) + 
\Gamma_{[k}\,^i\,_{|m|} \Gamma_{l]}\,^m\,_{j}
- \Gamma_{[k}\,^m\,_{l]}\,\Gamma_m\,^i\,_j = 1/2\,\,C^i\,_{jkl},
\end{equation}
and the Bianchi equation
\begin{equation}
\label{bianchi}
\nabla_{i}\,C^{i}\,_{jkl} = 0,
\end{equation}
which must be satisfied by the conformal Weyl tensor of a vacuum solution. 

\vspace{.3cm}

The coordinates and the frame field have to be restricted by gauge conditions to obtain a useful PDE problem. For simplicity we assume in this chapter  the normal to ${\cal S}$ to be tangential  to ${\cal T}$ (the general analysis is found in \cite{friedrich:nagy}). We focus our attention to some neighbourhood of a given point $p \in \Sigma$. A type of gauge as described below will be referred to as {\it admissible}

It holds $x^0 = 0$ on ${\cal S}$  and $x^1$, $x^2$, $x^3$ are local coordinates near $p$ with $x^3 \ge 0$ and $x^3 = 0$ on $\partial {\cal S} = \Sigma$. 
The fields $e_A$, $A = 1, 2$,  are tangential to 
${\cal S}_c = {\cal S} \cap \{x^3 = c =const. \ge 0\}$.

On ${\cal M} \sim [0, T[ \times {\cal S}$ the time-like unit vector field $e_0$ is future directed, tangential to ${\cal T} = [0, T[ \times \partial {\cal S}$, and orthogonal to  $S_c$.
The coordinate $x^0$ is a natural parameter on the integral curves of $e_0$ and the other coordinates are dragged along with $e_0$ so that $e_0(x^{\mu}) = \delta^{\mu}\,_0$.
The fields $e_A$ are $D$-Fermi-transported in direction of $e_0$ on
${\cal T}_c = \{x^3 = c\}$, where $D$ denotes the covariant derivative defined by the metric induced on ${\cal T}_c$. The field $e_3$ is normal to ${\cal T}_c$ and inward pointing on 
${\cal T} = {\cal T}_0$.

In this gauge we have
\[
e^{\mu}\,_0 = \delta^{\mu}\,_0, \quad 
D_a\,e_b = \Gamma_a\,^c\,_b\,e_c, \quad
\chi_{ab} \equiv \Gamma_a\,^3\,_b = \Gamma_b\,^3\,_a, \quad
\chi \equiv g^{ab}\,\chi_{ab}, \quad a, b, c = 0, 1, 2,
\]
with $\chi_{ab}$ the second fundamental form 
and $\chi$ the mean extrinsic curvature on ${\cal T}_c$, 
\begin{equation}
\label{e0-eA-transport}
\Gamma_0\,^A\,_B = 0, \quad 
D_{e_0}\,e_0 = \Gamma_0\,^A\,_0\,e_A, \quad
D_{e_0}\,e_A = - g_{AB}\,\Gamma_0\,^B\,_0\,e_0, \quad A, B = 1, 2,
\end{equation}
where the summation convention applies to both groups of indices.

As discussed in \cite{friedrich:nagy}, the three functions $\Gamma_0\,^A\,_B$, $\Gamma_0\,^A\,_0$ play the role of gauge source functions on ${\cal M}$ (cf. \cite{friedrich:2hyp red}) and can be chosen arbitrarily. The first of these functions has been disposed of here in a convenient way, choosing the other two functions represents the usual gauge problem in the  interior of ${\cal M}$ but it is delicate task on the boundary. The function $\chi$, which can also be prescribed, plays the role of a gauge source function in the interior of ${\cal M}$ and the role of a boundary datum on ${\cal T}$.

\vspace{.2cm}

With this gauge it is easy to  extracted from the complete, overdetermined system (\ref{first-structural}), (\ref{second-structural}), (\ref{bianchi}) a `reduced system' for the unknowns (that are not gauge source functions) which is symmetric hyperbolic and for which 
well-posed initial boundary value problems can be formulated. But in general the resulting evolution will not preserve the constraints so that not all of equations (\ref{first-structural}), (\ref{second-structural}), (\ref{bianchi}) will be satisfied.

The main theorem of \cite{friedrich:nagy} shows, however, that there do exist `reduced systems' for which well-posed initial boundary value problems can be formulated  whose solutions do satisfy the complete system (\ref{first-structural}), (\ref{second-structural}), (\ref{bianchi}). The following discussion refers to this result. 

It is well understood how to provide standard Cauchy data for Einstein's vacuum field equations on the space-like initial hypersurface ${\cal S}$. The  Weyl curvature is then derived from these data ${\cal S}$ by using the Gauss-Codazzi equations. We shall not repeat the details here and concentrate instead on the structure of the boundary conditions and the boundary data on ${\cal T}$, which are critical for our discussion.

It is convenient to  use on ${\cal T}$ a double null frame $l$, $k$, $m$, $\bar{m}$ so that, with $i^2 = - 1$,
\[
l = \frac{1}{\sqrt{2}}(e_0 + e_3),\quad
k = \frac{1}{\sqrt{2}}(e_0 - e_3),\quad
m = \frac{1}{\sqrt{2}}(e_1 + i\,e_2).
\]
Since the fields $e_1$, $e_2$ have been fixed only up to rotations $m \rightarrow e^{i\,\phi}\,m$ with functions $\phi \in C^{\infty}({\Sigma}, \mathbb{R})$, there is a corresponding freedom in the data. The `spin-weights' 
given below refer to the phase factors picked up under these rotations by the various quantities. 

\vspace{.1cm}

\noindent
In  \cite{friedrich:nagy} the following data are prescribed on ${\cal T}$: 

\vspace{.1cm}

\noindent
$-$ The mean extrinsic curvature $\chi \in C^{\infty}({\cal T}, \mathbb{R})$,

\vspace{.1cm}

\noindent
$-$ functions $q, \, \alpha, \, \beta, \, \Gamma \equiv  \Gamma_0\,^1\,_0 + i\, \Gamma_0\,^2\,_0
\in C^{\infty}({\cal T}, \mathbb{C})$ which are of
spin-weight $-2, -4, 0, 1$\\
\hspace*{.4cm}respectively and so that 
\[
\left[
\begin{array}{cc}
Re(\bar{\alpha}\,\beta) - \frac{1}{2}(1 - |\alpha|^2 - |\beta|^2)
& Im(\bar{\alpha}\,\beta) \\
Im(\bar{\alpha}\,\beta) & 
- Re(\bar{\alpha}\,\beta) - \frac{1}{2}(1 - |\alpha|^2 - |\beta|^2)
\\
\end{array}
\right] \le 0,
\]
\hspace*{.4cm}in the sense of quadratic forms.

\vspace{.2cm}

\noindent
The following boundary condition is required  on  ${\cal T}$:
\begin{equation}
\label{bdry-cond-*}
q = - \Psi_4 + \alpha\,\Psi_0 +
\beta\,\bar{\Psi}_0,
\end{equation}
where
\[
\quad\mbox{with}\quad
\Psi_0 = C_{\mu \nu \lambda \rho}
\,l^{\mu}\,m^{\nu}\,l^{\lambda}\,m^{\rho},\,\,\,\,\,\,\,\,
\Psi_4 = C_{\mu \nu \lambda \rho}
\,\bar{m}^{\mu}\,k^{\nu}\,\bar{m}^{\lambda}\,k^{\rho}.
\]

\subsection{Properties of the setting and the main problem}

If the gauge source functions and the functions $\alpha$, $\beta$ have been prescribed, the symmetric hyperbolic reduced system and the Cauchy data allows us to determine for all unknowns 
their time derivatives at of any order  on ${\cal S}$. The boundary condition 
(\ref{bdry-cond-*}) then gives us  restrictions on the function $q$ on $\Sigma$.  If $q$ is chosen accordingly, the consistency conditions will be satisfied.

Because the gauge conditions of the present setting are explicit, an analogue of 
(\ref{gamma=0-on-T}) is not required. Instead of controlling the gauge condition one has to control constraints. This task more or less motivated the way the problem has 
been arranged in \cite{friedrich:nagy}.

The functions $\alpha$ and $\beta$  do not carry critical information but they allow us to obtain physically or geometrically convenient formulations of the problem. As an example we note that  
special choices of $\alpha$ and $\beta$ lead to expressions for (\ref{bdry-cond-*}) which only involve 
certain components of either the electric  or the magnetic part of the conformal Weyl tensor with respect to $e_0$ or $e_3$.

The freedom which remains in the definition of the admissible gauge would allow us to choose  $\Gamma = 0$ near $\Sigma$. This implies, however, that the field $e_0$ is geodesic with respect to the metric  induced on ${\cal T}$,  that its flow lines may develop caustics, and that the gauge may break down after some finite time. Finding good choices of $\Gamma$  appears to correspond to the usual gauge problem in long term evolutions which also occur in the Cauchy problem. As seen below, there is much more to it.

A priori it cannot be excluced that parts of the boundary converge towards each other with a tendency to form selfintersections so that the boundary would stop being diffeomorphic to $\mathbb{R}_0^+ \times \Sigma$. One would try to avoid this by choosing  $\chi$ suitably. This is different from a change of gauge because it may  lead to changes of the geometry. It is not only that the manifold underlying the metric may change its  `size'  because the boundary evolves in a different way but the metric  itself may change in essential ways because the function $q$, given now on a different boundary hypersurface,
will acquire a new meaning with respect to the curvature.

It is, of course, understood that the `free' functions $\chi$ and $\Gamma_0\,^A\,_0$ extend smoothly into the interior of ${\cal M}$ as gauge source functions. The question whether the solution is independent of the choice of extension has been discussed in detail in \cite{friedrich:nagy}. A certain problem was left open there but I expect that it can be resolved. 

Irrespective of this problem the results of \cite{friedrich:nagy} guarantee the existence of (unique) solutions  locally near ${\cal S}$  for given 
initial-boundary data. In particular, given an $ST$--vacuum-space-time in one of the standard gauges, we can prescribe functions $\alpha$ and $\beta$ on ${\cal T}$ with the appropriate spin weights and read off the boundary datum $q$ satisfying (\ref{bdry-cond-*}). Given now the implied boundary  data  on ${\cal T}$ and the Cauchy data on ${\cal S}$, the solution can be reconstructed uniquely by solving the field  equations. 

Let some  $ST$--vacuum-space-time $({\cal M}, g)$ be given in two different admissible gauges which coincide on ${\cal S}$ but which are such that their respective  time-like unit vector fields $e_0$ and $e'_0$ do not coincide on 
${\cal T}$ in the future of $\Sigma$. Corresponding to the two gauges choose functions $\alpha$,  $\beta$ and $\alpha'$, $\beta'$  respectively in (\ref{bdry-cond-*}).
If the initial data induced on ${\cal S}$ and 
the two sets of data $(\chi, q, \Gamma)$ and $(\chi', q', \Gamma')$
 induced on ${\cal T}$ in are read off in these different gauges 
and the corresponding gauge source functions are use together with
the field equations to reconstruct the space-times, we can conclude by
PDE uniqueness that the two resulting solutions are isometric because we  know them to represent $({\cal M}, g)$ in two different gauges.

\vspace{.1cm}

\noindent
We can now state our main question:

{\it What could be said about the relation between the solutions determined by the two sets of initial boundary data without having the information that the data  have been derived from the same solution by employing two different admissible gauges ?}

\vspace{.1cm}

Given two sets of Cauchy data one can ask the analogous question whether the corresponding maximal, globally hyperbolic developments in time admit an isometry which maps the respective emdedded data hypersurfaces onto each other. 
This may be quite difficult to decide in a concrete case but in principle there exists a clear criterion which only involves the two Cauchy data sets: they must be isometric. Unfortunately, such a straightforward  answer seems not to be available  in the general situation considered above. 

Both the functions $\chi$ and $\chi'$ represent the mean extrinsic curvature 
of the hypersurface ${\cal T}$.
They must be given, however,  in the form $\chi = \chi(x^{\alpha})$ respectively $\chi' = \chi'(x^{\alpha'})$ with coordinates $x^{\alpha}$ and
$x^{\alpha'}$, $\alpha, \alpha'  = 0, 1, 2$, which are different from each other 
on ${\cal T} \setminus \Sigma$ because $e_0 \neq e'_0$ there. To see that $\chi$ and $\chi'$  represent the same object the coordinate transformation $x^{\alpha} = x^{\alpha}(x^{\beta'})$ needs to be known. 
One might think of deriving this transformation by using the data $\Gamma$, $\Gamma'$  in the  last two equations of  (\ref{e0-eA-transport}). But the boundary data do not contain any information on the metric induced on ${\cal T}$. The 
operators $D$ and $D'$ are thus only available after the full solution has been determined  on ${\cal T}$. 

The functions $q$ and $q'$ are difficult to compare for still another reason.  The frames
$e_k$, $e'_j$ corresponding to the two gauges are related  on ${\cal T}$ 
by a point dependent Lorentz transformation which leaves $e_3$ invariant. In Newman-Penrose notation one finds that the the components $\Psi_l$ and $\Psi'_k$, 
of the conformal Weyl tensor in the two frames are related by a transformation
\[
\Psi_k  \rightarrow \Psi'_k = \Psi_l\,s^l\,_k
\quad \mbox{with} \quad s^l\,_k \neq 0,\,\,\,\,l, k = 0, \ldots, 4.
\]
This  implies a relation
\[
q' = - \Psi'_4 + \alpha'\,\Psi'_0 + \beta'\,\bar{\Psi}'_0
= \eta\,q + \xi\,\bar{q} + \sum_{k = 0}^3\eta_k\,\Psi_k 
+ \sum_{k = 0}^3 \xi_k\,\bar{\Psi}_k
\]
with coeffficients $\eta_k$, $\xi_k$ which do not all vanish. The calculation shows that one cannot have $\eta_2 = 0$, $\eta_3 = 0$ simultaneously near $\Sigma$ in 
${\cal T} \setminus \Sigma$ if $e'_0 \neq e_0$ there. The relation above thus involves components of the conformal Weyl tensor which are not provided by the given boundary data and which will become available only after the development of the solutions in time has been determined.

This suggests that {\it under general assumption there does not exist a reasonable concept of a
`diffeomorphism class of initial boundary data'}. The boundary data sets $(\chi, q, \Gamma)$ and $(\chi', q', \Gamma')$
cannot be compared by operations on ${\cal T}$ which only involve these data sets and possibly interior equations induced by the field equations  on ${\cal T}$. The comparison requires  knowledge of  the development in time of the two initial boundary data sets.

It should be emphasized  that rotations of the frame $e_A$
tangential to ${\cal S}_c$ are reflected in the boundary data by simple phase transformations. The problems of the present approach arise from the need to single out a time-like vector field tangential to the boundary ${\cal T}$. In general, there does not exist a distinguished choice.

\vspace{.1cm}

While this is also true in the case of solutions of Anti-de Sitter type studied in \cite{friedrich:AdS}, geometric uniqueness can be shown. It follows because the initial boundary data are given by standard Cauchy data on ${\cal S}$ and a Lorentzian conformal structure on the conformal boundary ${\cal J}$ which satisfy  the appropriate consistency conditions. The possibility to arrive at such a geometric statement, which has no need for a time-like vector field, is related to the fact that the boundary is 
not  simply `put in by hand' but a consequence of  the compatibility of the field equations with the requirement  of the existence of a conformal boundary ${\cal J}$.

As a consequence, the geometric fields induced on the boundary have the following special properties.
There exists a gauge based on the conformal structure of $g$ which reduces on ${\cal J}$ to an analogous gauge based on the induced conformal structure on 
${\cal J}$. With a suitable choice of $\alpha$ and $\beta$ a certain component of
the `boundary-magnetic part'    of the rescaled Weyl tensor represents the free datum
$q$.
This component  is related to the Coton tensor of the intrinsic conformal
structure on ${\cal J}$. Via this relation it gives precisely the information needed in
the structural equations of the normal conformal Cartan connection
to determine from $q$ the conformal structure induced on ${\cal J}$ and vice versa.

\section{An approach based on wave equations}

To see how the features pointed out above are reflected in a different formulation, we give an outline of the approach of \cite{kreiss:et:al:2009}, in which Einstein's vacuum field equations are understood as equations of second order for the metric coefficients. 

\subsection{Setting up the PDE problem}

The authors of \cite{kreiss:et:al:2009} consider a fixed background metric 
$\hat{g}_{\mu \nu}$ on ${\cal M} = [0, t_*[ \times {\cal S}$ with some $t_* > 0$ so that the slices 
$\{t\} \times {\cal S}$, with $0 \le t < t_*$, ${\cal S} \equiv \{0\} \times {\cal S}$,  are space-like and the boundary 
${\cal T} = [0, t_*[ \times \partial {\cal S}$ is time-like with respect to $\hat{g}_{\mu \nu}$.
They impose on ${\cal M}$  the gauge condition
\begin{equation} 
\label{non-harm-map}
0 = C^{\mu} \equiv g^{\lambda \rho}\,(\Gamma_{\lambda}\,^{\mu}\,_{\rho}
- \hat{\Gamma}_{\lambda}\,^{\mu}\,_{\rho}) - H^{\mu}
= g^{\lambda \rho}\,g^{\mu \nu}\,(
\hat{\nabla}_{\lambda}\,h_{\rho\nu} 
- 1/2\,\,\hat{\nabla}_{\nu}\,h_{\lambda \rho}
) - H^{\mu},
\end{equation}
where $H^{\mu}$ denotes a given vector field, 
$\Gamma_{\lambda}\,^{\mu}\,_{\rho}$, $\hat{\Gamma}_{\lambda}\,^{\mu}\,_{\rho}$ denote
the Christoffel symbols of $g_{\mu \nu}$, $\hat{g}_{\mu \nu}$ respectively, 
$\nabla$, $\hat{\nabla}$ denote the covariant operators defined by $g$ and $\hat{g}$,
and
\[
h_{\mu \nu} = g_{\mu \nu} - \hat{g}_{\mu \nu}.
\]
In this gauge the reduced vacuum field equations take the form of a system of wave equations 
\begin{equation}
\label{red-wave-equ}
 g^{\lambda \rho}\,\hat{\nabla}_{\lambda}\,\hat{\nabla}_{\rho}\,h_{\mu \nu} 
= F_{\mu \nu}(h,\, \hat{\nabla}\,h,\,g,\,Riem[\hat{g}]) 
+ \nabla_{(\mu}\,H_{\nu)}, 
\end{equation}
with a certain polynomial function $F_{\mu \nu}$. 

\vspace{.1cm}

Boundary conditions are imposed as follows. With respect to g let $T$ denote a future directed time-like unit vector field tangential to ${\cal T}$, $N$ the outward directed unit normal to ${\cal T}$, and define the real null vectors  
$K = T + N,\quad  L = T - N$. Let $Q$ be a complex linear combination of vectors orthogonal to $K$ and $L$ which satisfies
together with its complex conjugate $\bar{Q}$ the normalization
conditions $g(Q, Q) = 0$, $g(Q, \bar{Q}) = 2$.
The metric can then be written
\begin{equation}
\label{K-L-Q-metric-expr}
g_{\mu \nu} = 1/2\,\left(Q_{\mu}\,\bar{Q}_{\nu} + \bar{Q}_{\mu}\,Q_{\nu} 
- K_{\mu}\,L_{\nu} - L_{\mu}\,K_{\nu}\right).
\end{equation}
With suitable  functions 
\[
q_{KK}, \, \, q_{Q\bar{Q}} \in C^{\infty}({\cal T}, \mathbb{R}), \quad \quad
q_{KQ}, \,\, q_{QQ} \in C^{\infty}({\cal T}, \mathbb{C}),
\]
as boundary data, the gauge requirements
\[
\left(K^{\mu}\,C_{\mu}\right)_{\cal T} = 0, \quad 
\left(L^{\mu}\,C_{\mu}\right)_{\cal T} = 0, \quad
\left(Q^{\mu}\,C_{\mu}\right)_{\cal T} = 0, 
\]
on ${\cal T}$, where the metric on the right hand side of (\ref{non-harm-map}) is written as in  (\ref{K-L-Q-metric-expr}),
are supplemented by the boundary conditions
\begin{equation}
\label{a-bdry-con}
\left(K^{\mu}\,K^{\nu}\,K^{\rho}\,\hat{\nabla}_{\mu}\,h_{\nu \rho}
+ \frac{2}{r}\,K^{\nu}\,K^{\rho}\,h_{\nu \rho}\right)_{\cal T} = - q_{KK},
\end{equation}
\begin{equation}
\label{b-bdry-con}
\left(K^{\mu}\,K^{\nu}\,L^{\rho}\,\hat{\nabla}_{\mu}\,h_{\nu \rho}
+ \frac{1}{r}\,K^{\nu}\,L^{\rho}\,h_{\nu \rho}
+ \frac{1}{r}\,Q^{\nu}\,\bar{Q}^{\rho}\,h_{\nu \rho}\right)_{\cal T}
= - q_{Q\bar{Q}},
\end{equation}
\begin{equation}
\label{c-bdry-con}
\left(K^{\mu}\,K^{\nu}\,Q^{\rho}\,\hat{\nabla}_{\mu}\,h_{\nu \rho}
+ \frac{2}{r}\,K^{\nu}\,Q^{\rho}\,h_{\nu \rho}\right)_{\cal T} = - q_{KQ},
\end{equation}
\begin{equation}
\label{d-bdry-con}
\left(K^{\mu}\,Q^{\nu}\,Q^{\rho}\,\hat{\nabla}_{\mu}\,h_{\nu \rho}
- Q^{\mu}\,Q^{\nu}\,K^{\rho}\,\hat{\nabla}_{\mu}\,h_{\nu \rho}\right)_{\cal T}
= - q_{QQ},
\end{equation}
where the function $r$ denotes the areal radius of the cross section $\{ t \} \times \partial S$ with
respect to the background metric.

\vspace{.1cm}

\subsection{Some features of the PDE problem}

No detailed specification of the frame $K$, $L$, $Q$, $\bar{Q}$ has been made in
\cite{kreiss:et:al:2009}. To obtain a well defined PDE problem one has to decide, however, on a definit prescription. This choice enters the consistency conditions, it determines the evolution of the boundary and the gauge, and it  affects the isometry class of the solution if the right hand sides of 
(\ref{a-bdry-con}) - (\ref{d-bdry-con}) are given.

Since the metric is not available on ${\cal T}$ at this stage, one has to
give an abstract prescription for the frame field. Assuming $\partial_0$ to be time-like
and $\partial_a$, $a = 1, 2, 3$, to be space-like on $\Sigma = {\cal S} \cap {\cal T}$,
the specification of the frame can then be made in terms of a Gram-Schmidt
orthonormalization of the coordinate frame, which gives the frame  $K$, $L$, $Q$,
$\bar{Q}$ in terms of the yet unknown metric coefficients $g_{\mu \nu}$ near $\Sigma$. Note that
there is a large freedom in doing this and that it will not be clear a priori for how long into the future a particular choice will be well behaved. 

Again it is the choice of the time-like unit vector field $T$ which is decisive here.  
The dependence of the boundary data on the complex vector fields $Q$ and $\bar{Q}$ orthogonal to $T$ and $N$  is well controlled because the boundary data  pick up phase factors $e^{i\, k \,\psi}$, with certain $k \in \mathbb{Z}$,
under rotations  $Q^{\mu} \rightarrow e^{i\,\psi} \,Q^{\mu}$ with $\psi \in C^{\infty}({\cal T}, \mathbb{R})$. In the following a definite
prescription of the frame will be assumed.

The meaning of the boundary conditions has been discussed to some extent in
\cite{kreiss:et:al:2009}, but it is not easy to analyse and probably not fully understood yet. They affect in particular the gauge defined by  (\ref{non-harm-map}).
The tensorial nature  of this gauge condition allows one to change the coordinates conveniently without affecting the hyperbolicity of the reduced equations.
The fields $H^{\mu}$ reflect the usual freedom to prescribe four gauge source functions. Their use allows us to generalize the maps of $({\cal M}, g)$ onto $({\cal M}, \hat{g})$ considered below
to {\it general wave maps}. 

To simplify the discussion we consider here only the case where $H^{\mu} = 0$. 
If $g$ exists on ${\cal M}$, relation  (\ref{non-harm-map}) then tells us that 
the identity map of ${\cal M}$ defines a {\it wave map} of $({\cal M}, g)$ onto $({\cal M}, \hat{g})$. More generally, a map $\Phi: {\cal M} \rightarrow {\cal M}$ is a wave map  for $g$ and $\hat{g}$ if it satisfies the variational principle $\delta \int tr_{g}(\Phi^* \hat{g})\,d\mu_g = 0$. Let $(U, x^{\mu'})$ denote a coordinate patch on ${\cal M}$, $x^{\mu}$ coordinates defined on $\Phi (U)$, and $\Phi^{\mu}(x^{\mu'})$ the local representation of $\Phi$. The Euler-Lagrange equations of this principle are then equivalent to
the system of wave equations 
\begin{equation}
\label{harm-map-equ}
g^{\mu' \nu'}\left(\Phi^{\rho}\,_{, \mu' \nu'} - \Gamma_{\mu'}\,^{\lambda'}\,_{\nu'}\,\Phi^{\rho}\,_{, \lambda'}
+ \hat{\Gamma}_{\mu}\,^{\rho}\,_{\nu}\,\Phi^{\mu}\,_{, \mu'}\,\Phi^{\nu}\,_{, \nu'} 
\right) = 0,
\end{equation}
where $ \Gamma_{\mu'}\,^{\lambda'}\,_{\nu'}$ and $\hat{\Gamma}_{\mu}\,^{\rho}\,_{\nu}$
denote the Christoffel symbols of the metrics $g_{\mu' \nu'}$ and $\hat{g}_{\mu \nu}$ respectively
and $\hat{\Gamma}_{\mu}\,^{\rho}\,_{\nu}$ is taken at $\Phi^{\mu}(x^{\mu'})$.

Such maps may be constructed by  solving initial boundary value problems for (\ref{harm-map-equ}) with  Cauchy data on ${\cal S}$ and boundary conditions resp. data on ${\cal T}$. To allow them to define diffeomorphisms ${\cal M} \rightarrow {\cal M}$ of the desired type, the  data should given such that the
maps  induce diffeomorphisms of ${\cal S}$ and ${\cal T}$ onto themselves respectively and such that the tangent  maps $T_q \Phi$
have  maximal rank at points $q$ in  ${\cal S}$ or  ${\cal T}$. A solution will then map some neighbourhood of ${\cal S}$ in ${\cal M}$ diffeomorphically onto another such neighbourhood.
Depending on the prescribed data and the metrics $g$ and $\hat{g}$, 
its life time as a diffeomorphism may be limited, however.

Assume such a map $\Phi$ to be given, denote its inverse by $\Psi$  and the pull back of $g$ under $\Psi$ by $g' = \Psi^*g$. Writing  the left hand side of  (\ref{harm-map-equ})  in terms of the argument  $x^{\mu}$, one finds that the equations  can be rewritten in the form 
\begin{equation}
\label{harm-map-transf}
g'^{\lambda \rho}\,(\Gamma'_{\lambda}\,^{\mu}\,_{\rho}
- \hat{\Gamma}_{\lambda}\,^{\mu}\,_{\rho}) = 0.
\end{equation}
This is (\ref{non-harm-map})  with $g$ and its Christoffel symbols replaced by $g'$
and its derived fields. Condition (\ref{non-harm-map}) thus remains unchanged if we allow for pull backs of $g$ by  inverses of wave maps. This freedom is removed by the initial boundary conditions for 
(\ref{harm-map-equ}) which are implicit in the initial boundary conditions for (\ref{red-wave-equ}).

It is not easy to see whether the boundary conditions 
(\ref{a-bdry-con}) - (\ref{d-bdry-con})
contain information which can be related directly to the behaviour of the geometry defined by $g$ near the boundary. We shall ignore that point here,  though the question becomes important when the geometry 
develops a tendency to collapse near the boundary.

On the boundary  ${\cal T}$ we have the freedom to choose the time-like vector field $T$ and the related frame and the data  $q = (q_{KK}, \, q_{Q\bar{Q}}, \,q_{KQ}, \,q_{QQ})$. The question arises what happens under a transition 
\[
(T, q) \rightarrow (T', q') \quad \mbox{with} \quad T \neq T'. 
\]
It could be that it is just implied by a change of the gauge and the frame but it could also correspond to a transition to a different isometry class of metrics. Again it turns out to be impossible  to decide this only on the basis of the data given on ${\cal T}$. 

The boundary conditions (\ref{a-bdry-con}) - (\ref{d-bdry-con}) are covariant with respect to coordinate transformations but they have a complicated behaviour under gauge transformations. It is natural to accompany the  gauge transformation $g \rightarrow \Psi^* g$ 
leading to (\ref{harm-map-transf}) by the push forward of the frame based on the harmonic map $\Phi$.  The defining properties of the frame will be preserved and some calculations will be simplified. 
The transformation laws of the functions
comprised by $q$ follow from the transformation laws  of the left hand sides of the boundary conditions. Because the background metric 
$\hat{g}$ and thus the operatore $\hat{\nabla}$ is kept fixed, the field $h = g - \hat{g}$ transforms into $\Psi^* g - \hat{g}$
and the transformation of $q$ will  involve derivatives of $\Psi$ up to second order.
Without knowing  the transformed solution near ${\cal T}$ these derivatives cannot be determined on ${\cal T}$. 

Even if the transition$(T, q) \rightarrow (T', q')$ above would result from a simple  redefinition of the frame we would not be able to recognize that. The transformation formula for $q$ under transformations of the frame which leave the normal vector $N$ fixed  requires information on the unkowns which is only  available when the solutions are known near ${\cal T}$.

We conclude that in both approaches, \cite{friedrich:nagy} and \cite{kreiss:et:al:2009}, 
the problem with geometric uniqueness is related to the frame dependence of the boundary data. There is an `{\it inner frame dependence}', which refers, depending on the method, to the coordinates in which the boundary data are given or to the inverse wave map acting on the metric, and an `{\it outer frame dependence}' which refers to the need to perform in a transformation  linear combinations of some of the given data with other data which are not available. In both cases it is the choice of the  time-like vector field tangential to ${\cal T}$ which is critical.  In general there does not exist a distinguished one.
Because it  is based on a very general analysis of initial boundary value problems for systems of wave equations, the approach of \cite{kreiss:et:al:2009} and of previous articles by the same authors may offer more flexibility than that of \cite{friedrich:nagy} and it may offer new and unexpected possibilities to address the problem of geometric uniqueness.

\section{Covariant boundary data and distinguished time-like vector fields}

To avoid problems arising from the frame dependence one may wish to find formulations in which the boundary data which do not serve to control the gauge are prescribed in terms of  frame independent fields. In an approach which employs gauge conditions such as (\ref{gamma=0-on-T}) and uses wave equations as reduced equations they should be given in terms of the first and second fundamental form
\[
k_{\mu \nu} = g_{\mu \nu} - N_{\mu}\,N_{\nu}, \quad 
\chi_{\mu \nu} = k_{\mu}\,^{\lambda}\,k_{\nu}\,^{\rho}\,\nabla_{\lambda}\,N_{\rho},
\]
induced on ${\cal T}$. Here $N$ denotes the outward pointing unit normal to 
${\cal T}$ with respect to $g$. The first invariant associated with these fields in the mean extrinsic curvature  $\chi = k^{\mu \nu}\,\chi_{\mu \nu}$. Two further invariants are supplied by the eigenvalues of the trace free part
\[
\chi^*_{\mu \nu} = \chi_{\mu \nu} - \frac{\chi}{3}\,k_{\mu \nu},
\]
of the second fundamental form. In local coordinates $x^{\alpha}$ 
on ${\cal T}$ in which the induced metric takes at a given point $p' \in {\cal T}$ the standard form
$k_{\alpha \beta} = \eta_{\alpha \beta} \equiv diag(-1, 1, 1)$ the 
field $\chi^{* \mu}\,_{\nu} = k^{\mu \rho}\,\chi^*_{\rho \nu} $ is represented at $p'$, possibly after a rotation of the coordinates which leaves the $x^0$-axis fixed, by a matrix of the form
\[
\chi^{*\alpha}\,_{\beta} =
 \left(\begin{array}{ccc}
b + d & - c^1 & - c^2\\
c^1 & - b & 0\\
c^2 & 0 & - d
\end{array}\right).
\]
The eigenvalues $\lambda_i$, $i = 0, 1, 2$, of $\chi^*_{\alpha \beta}$, which satisfy $\lambda_0 + \lambda_1 + \lambda_2 = 0$, are then the roots of the equation
\[
0 = \det(\chi^{*\alpha}\,_{\beta} - \lambda\,\delta^{\alpha}\,_{\beta}) = - \lambda^3  
 + 1/2\,\chi^{*\alpha}\,_{\beta}\,\chi^{*\beta}\,_{\alpha}\,\lambda + \det(\chi^{*\alpha}\,_{\beta}).
\]
They are functions of  the coefficients of this equation and one might try to use directly the invariants
\begin{equation}
\label{chi-invariants}
\chi, \quad \,\, \chi^{*\alpha}\,_{\beta}\,\chi^{*\beta}\,_{\alpha}, \quad \,\,  \det(\chi^{*\alpha}\,_{\beta}),
 \end{equation}
as boundary data. Even if they could be used for that purpose there remains the problem that the way  they must be given may depend on the time-like vector field $T$. 

One of the eigenvalues of $\chi^{*\alpha}\,_{\beta}$ is necessarily real and we may ask whether  the real eigenvectors of $\chi^{*\alpha}\,_{\beta}$ can be of any use for us. If $b = d = 0$ and $c^A \neq 0$ two of the eigenvalues are complex conjugates of each other and there is one  real eigenvector which turns out to be space-like. This remains true if  
 $|c| = \sqrt{(c^1)^2 + (c^2)^2}$ is much larger than $|b|$ and $|d|$. If $c^A = 0$ there exists an orthonormal frame of eigenvectors of $\chi^{*\alpha}\,_{\beta}$ but the  time-like eigenvector need not be unique. In fact, if $b = - 2\,d$ or $d = - 2\,b$ there exists
a 2-dimensional, time-like subspace of eigenvectors.  

A view at the space-time setting underlying  the flat, linear model problem considered in the beginning may suggest  a reasonable condition under which $\chi^{*\alpha}\,_{\beta}$ can be expected to admit a unique time-like eigenvector.  In that case we find $\chi = 2/R > 0$ and $\chi^{*\alpha}\,_{\beta} = \frac{1}{R}\,(\frac{1}{3}\,k_{\mu \nu} + T_{* \mu}\,T_{*\nu})$.
It follows that $\chi^{*\alpha}\,_{\beta}$ admits  $T_* = \partial_t$  as its unique future directed, time-like unit eigenvector. The property that $T_*$ is orthogonal to the hypersurfaces $\{t = const.\}$ cannot be expected to extend to the curved case but the fact that the pull-back of $\chi^*_{\alpha \beta}$ to the plane orthogonal to $T_*$ is (positive) definite suggests a useful generalization.

Back to the general case, suppose that $\chi^{*\alpha}\,_{ \beta}$ has a time-like eigenvector 
$T \neq 0$ tangential to ${\cal T}$ at $p'$ with eigenvalue $\lambda_0$.
With a suitable scaling of $T$ we can assume, possibly after  
a Lorentz-Transformation,  which leaves the form $k_{\alpha \beta} = \eta_{\alpha \beta}$
unchanged, that $T^{\alpha} = \delta^{\alpha}_0$. It follows that
\[
\chi^*_{\alpha0}  = - \lambda_0\,\delta^0_{\alpha}.
\]
By a rotation about the $x^0$-axis, which leaves  the form of $k_{\alpha \beta}$ and
$T^{\alpha}$ unchanged, the symmetric trace free tensor 
$ \chi^*_{\alpha \beta}$ can then be brought into the diagonal form
\[
\chi^*_{\alpha \beta} = (\lambda_1 + \lambda_2)\,\delta^0_{\alpha}\,\delta^0_{\beta}
+  \lambda_1\,\delta^1_{\alpha}\,\delta^1_{\beta}
+  \lambda_2\,\delta^2_{\alpha}\,\delta^2_{\beta}.
\]

Cases in which the time-like eigendirection is not unique
are excluded if we assume that the pull- back of $\chi^*_{\alpha \beta}$ to the
hyperplane orthogonal to $T$ is positive definite (negative definite, with corresponding changes below, might also be considered). The representation above then implies 

\vspace{.1cm}

\centerline{ $(*) \quad$ {\it the quadratic form
$\chi^*_{\alpha \beta}\,X^{\alpha}\,X^{\beta}$ on the tangent space $T_{p'} {\cal T}$  is positive
definite}.}

\vspace{.1cm}

\noindent
Conversely, consider  on $T_{p'} {\cal T}$ the functions
$h = k_{\alpha \beta}\,X^{\alpha}\,X^{\beta}$, $f = \chi^*_{\alpha \beta}\,X^{\alpha}\,X^{\beta}$, 
and the set $H = \{X \in T_p {\cal T}|\,\,h(X)= -1\}$. Assuming $(*)$, we conclude that 
$f(X) \rightarrow \infty$ on $H$ if the direction of $X$ approaches the null cone of $k$.  The restriction of $f$ to $H$ thus assumes a minimum at some point $X_* \in H$ and by Lagrange's method of underdetermined multipliers there exists a real constant 
$\lambda$ with 
\[
\chi^*_{\alpha \beta}\,X_*^{\beta} = \lambda \,k_{\alpha \beta}\,X_*^{\beta},
\]
so that  $X_*$ is a time-like eigenvector of $\chi^*_{\alpha \beta}$. Because the restriction of the form $\chi^*_{\alpha \beta}\,X^{\alpha}\,X^{\beta}$ to the
plane orthogonal to $X_*$ is positive definite, we are again  in the situation which led to 
$(*)$.

We complement  the assumption above by 

\vspace{.1cm}

\centerline{ $(**) \quad$ {\it the mean extrinsic curvature $\chi$  is positive}.}

\vspace{.1cm}

\noindent
Conditions $(*)$ and $(**)$ then imply that the set ${\cal M}$ is locally convex at
${\cal T}$ in the following sense. Suppose that $({\cal M'}, g')$ is a smooth extension of the space-time $({\cal M}, g)$ for which ${\cal T}$ is an interior hypersurface. Any geodesics $\gamma(\tau)$ in this extension
which is tangential to ${\cal T}$ at the point $\gamma(0) \in {\cal T}$ will only be tangential to ${\cal T}$ at first order and remain outside ${\cal M}$ for 
$0 < |\tau| << 1$ if its tangent vector 
$\gamma'(0)$ belongs to the set 
$\{X \in T_{ \gamma(0)}{\cal T}| \,k(X, X) \ge 0\}$ or is sufficiently close to it.

Assumptions $(*)$,  $(**)$ are quite natural if we want to pose an initial boundary value problem for an interior part of an
asymptotically flat solution whose boundary is sufficiently close to space-like infinity. They single out a future directed, time-like unit  vector field tangential to ${\cal T}$ which is distinguished by the geometry of the problem. The question about the dependence of the invariants (\ref{chi-invariants}) on the frame does not arise any longer. Moreover, conditions $(*)$,  $(**)$ are preserved under  small perturbations. If $\chi$ and the eigenvalues could be prescribed as boundary data the conditions could in fact be ensured during the development in time.

Whether  the invariants (\ref{chi-invariants}) can be used  to encode the two degrees of freedom of the gravitational fields is not obvious. As mentioned before, the right hand side of the boundary condition (\ref{bdry-cond-*}) can be expressed completely in terms  of the $e_3$-magnetic part of the conformal Weyl tensor. By the Codazzi equations the latter is  given by certain covariant derivatives of $\chi_{\alpha \beta}$ on ${\cal T}$ so that the function $q$ is related to $\chi$ and the eigenvalues. Because the covariant derivatives involve the connection defined by $k_{\alpha \beta}$ it is not easy to see that the information encoded  in $q$ can be extracted from  the $\lambda_i$ and $\chi$.   
It may be that the situation is more easily analysed in the setting of  \cite{kreiss:et:al:2009}.

There is a test which may shed some light on this question. Consider one of the two approaches above
and assume that  the initial and boundary data  are given such that the solutions will coincide for $0 \le t < \epsilon$ with the space-time setting of the flat, linear model case for some small $ \epsilon > 0$.
Assume for $t \ge \epsilon$ the free functions entering the gauge condition and the specification of the boundary evolution to be given such that the gauge and the boundary reduce to that of the flat model case if the boundary data which refer to the gravitational degrees of freedom are given such as to imply the setting of the flat model case. 
Consider now a solution which is determined by some given but unspecified  boundary data. We can ask then  whether  the additional requirement that the eigenvalues of $\chi_{\alpha \beta}$ coincide with those  of the flat model case implies  that the solution must be flat. 
A positive answer can be expected to indicate that the eigenvalues constitute suitable boundary data. Moreover, the argument which leads to this answer may give some insight into how initial boundary value problems which include these data must be formulated.

\vspace{.1cm}

There remains, of course, the complicated question whether a time-like eigenvector or even an eigenframe of the second fundamental form can be implemented together with a condition of the type
(\ref{gamma=0-on-T}) and possibly a prescription of the eigenvalues of the second fundamental form
in a formulation of a well posed initial boundary value problem. The answer requires a detailed analysis which will not be attempted here.

\section{Conclusions}

The formulations of the boundary conditions considered in \cite{friedrich:nagy} and 
\cite{kreiss:et:al:2009} require the choice of a future directed, time-like unit vector field $T$ tangential to the boundary. With the resulting initial boundary conditions  and data one arrives at well posed PDE problems
which determine $ST$-vacuum-space-times locally in time near  the initial hypersurface ${\cal S}$. The latter are  unique apart from possible extensions into the future. Moreover,  any  $ST$-vacuum-space-time which is given in one of the gauges considered above can be constructed in this way  locally in time.

The vector field $T$, for which no natural choice exists  in general,  is characterized indirectly  and becomes explicitly  available only after solving the equations. Problems arise if one wants to compare solutions pertaining to boundary conditions based on different choices of $T$ and on different boundary data.

The boundary conditions and data contain only very little direct  information on the geometry on the boundary  and the meaning of  the boundary data is related to the choice of $T$. As a consequence, the question whether   solutions determined by two different sets of boundary conditions and data are isometric can not be answered in terms of the boundary conditions and data alone. The complete solutions  must be available along the boundary to perform a comparison. 

This situation leads to awkward practical problems if gauge transformations need to be considered in the course of  an evolution. It is an open question whether this is an intrinsic problem of the initial boundary value problem for Einstein's field  equations  or whether  there can be formulated, under general assumptions,  initial boundary value problems for Einstein's field  equations which avoid these difficulties. 

We pointed out a class of initial boundary value problem for which a time-like vector  field $T$ is distinguished by the geometry of the boundary. Its defining property is stable under perturbations and the class is fairly large and quite important from the point of view of applications. If there existed formulations of well posed initial boundary problems based on this vector field and possibly on the invariants considered above
the problem with geometric uniqueness arising in  more general situations  would not be present.

\vspace{.3cm}

\noindent
{\bf Acknowledgement.} This article was inspired by seminars and discussions during the program  on `Geometry, Analysis, and General Relativity' at the
Institut Mittag-Leffler. I should like to thank the institute for hospitality, financial  support, and a most stimulating environment.

}

\end{document}